\documentclass[%
reprint,
superscriptaddress,
amsmath,amssymb,
prl,
nofootinbib,
nobibnotes
]{revtex4-1}

\usepackage{verbatim,xcolor,graphicx,mathtools,epstopdf,subfigure}
\usepackage[utf8]{inputenc}
\usepackage[UKenglish]{babel}

\begin{document}
   \title{Deal-Grove Oxidation and Nanoparticle Adhesion -- an AFM Study}
   
   \author{Clara C. Wanjura}\email{ccw45@cam.ac.uk}
   \affiliation{Ulm University, Albert-Einstein-Allee 11, 89081 Ulm, Germany}
   \affiliation{Current address: Cavendish Laboratory, Cambridge University, \\ JJ Thomson Avenue, Cambridge CB3 0HE, UK}
   \author{Daniel Geiger}
   \affiliation{Ulm University, Albert-Einstein-Allee 11, 89081 Ulm, Germany}
   \author{Irina Schrezenmeier}
   \affiliation{Ulm University, Albert-Einstein-Allee 11, 89081 Ulm, Germany}
   \affiliation{Current address: Carl Zeiss SMT GmbH, Rudolf-Eber-Straße 2, 73447 Oberkochen, Deutschland}
   \author{Matthias Roos}
   \affiliation{Carl Zeiss SMT GmbH, Rudolf-Eber-Straße 2, 73447 Oberkochen, Deutschland}
   \author{Othmar Marti}
   \affiliation{Ulm University, Albert-Einstein-Allee 11, 89081 Ulm, Germany}

   \date{\today}
   
   \begin{abstract}
      We study the formation of nanometer thick oxide layers around nanoparticles (plateaus) in a wet nitrogen environment over $32$ days. We determine both the plateau height and the lateral force needed to remove the particles with an atomic force microscope (AFM). The height is well described by the Deal-Grove model for a thermally activated diffusion limited oxidation process. Furthermore, we observe that height and force appear to be correlated suggesting that the rise in adhesion forces is mainly governed by an oxidation process.
      Since the plateaus are the result of a change in the chemical structure, the surface remains permanently altered even after the removal of the nanoparticles. This observation is of great importance for many applications requiring smooth silicon surfaces, such as photolithography, where small contamination can cause the observed oxidation process and significantly increase the surface roughness.
   \end{abstract}
   
   
   \maketitle
   
   \section{Introduction}
      Recent technological advances have made understanding the process by which nanoparticles adhere to surfaces of greatest importance \cite{bib:Carrillo,bib:Tian2017,bib:Tian2018,bib:Tian2018PhysChem}. Devising means to reduce the force with which nanoparticles stick to surfaces, is crucial in many applications. In photolithographic systems, for instance, small particles may lead to undesired light scattering causing intensity losses and interference which can destroy or damage the device geometry. In other fields the opposite, strong particle adhesion, may be desirable \cite{bib:Haun,bib:vdMaaden,bib:Lamprecht}.
      
      In our study, we focus on adhesion in the model system of spherical silica nanoparticles on a silicon surface, which is of particular interest in the aforementioned applications. 
      Frictional aging has been the focus of recent studies \cite{bib:Li,bib:Tian2017,bib:Tian2018,bib:Tian2018PhysChem} examining rate and state friction laws in slide-hold-slide experiments with an atomic force microscope.
      Multiple processes contributing to the adhesion have already been identified, ranging from a thin water film on the surface \cite{bib:opitz}, as well as the formation of a water meniscus \cite{bib:Kwon} between particles and surface, to electrostatic interaction or chemical reactions \cite{bib:ranade,bib:Li}.
      
      An important aspect regarding the adhesion mechanism we claim stems from the oxidation of the surface around the particles. Oxidation of silicon surfaces in general has been widely researched starting from Deal and Grove in 1965 \cite{bib:dealGrove} describing the thermally activated, diffusion limited oxidation. Oxide growth has been investigated experimentally in different temperature regimes \cite{bib:dealGrove,bib:Deal}, among other things at room temperature~\cite{bib:Morita}. They identified the presence of water and oxygen as crucial ingredients for the growth process under these conditions. The chemical components forming in the reaction have been analysed by Ling et al~\cite{bib:Ling} showing that not only oxides but also hydrogen compounds are created. The kinetics of contributing reactions were studied theoretically by Lora-Tamayo et al \cite{bib:Lora-Tamayo1978}. Furthermore, Gould et al showed that cleaning agents also have an impact on the oxide growth~\cite{bib:GouldIrene}. Applications of a controlled oxide growth in electrical insulators were researched by Ohmi et al \cite{bib:Ohmi}.
      
      While the subject of all of these studies was the growth of extended oxide layers, we report the local formation of oxide layers of nanometer thickness, which we dub plateaus, beneath and around nanoparticles in a wet nitrogen environment at room temperature. They also grow in ambient air. The result is a permanently damaged surface with an increased roughness.
      
      Using atomic force microscopy (AFM), we determine the plateau height as function of the storage time over $32$ days. We show that the growth height is compatible with the Deal-Grove model. In addition, we quantify the interactions between particle and surface by measuring the lateral force needed to remove a single particle with an AFM tip in contact mode. We demonstrate a correlation between lateral force and plateau thickness suggesting that the oxidation is a major contributor to the particle adhesion.
      
   \section{Experimental Procedure}
         {\bf\noindent Sample Preparation:} The samples are prepared on silicon wafers~\cite{bib:si-mat} with an epitaxial grown surface layer covered by their native oxide layer. They are coated with a diluted solution of $50\,\mathrm{nm}$~\cite{bib:siSolution} sized silica nanoparticles through spin coating. This yields samples with mainly solitary nanoparticles decorated on the surface ideal for individual manipulation and measurement.
      
         {\bf\noindent Measurements and Data Processing:} In order to obtain good statistics and high reproducibility for both the lateral force and the plateau height measurements, we use an automated procedure. The experiments are performed with Bruker's AFM Multimode V equipped with silica cantilevers~\cite{bib:brukerAFM}.
         \begin{figure}[htbp]
            \centering
            \includegraphics[width=.48\textwidth]{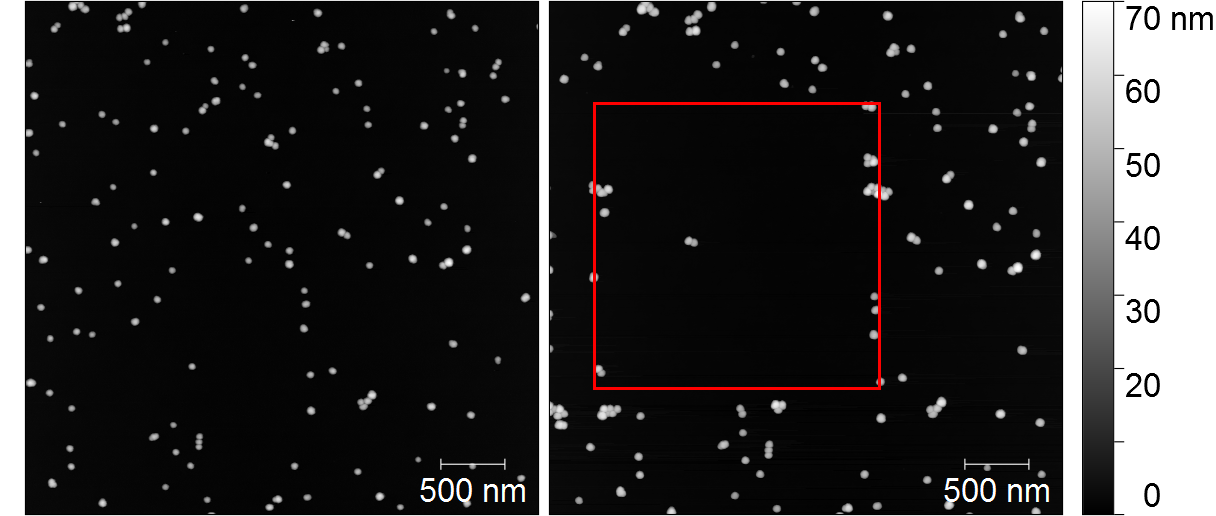}
            \caption{(Color online) Scanning region (red square) before and after the lateral force measurement: particles are clearly removed by the sweeping motion of the AFM tip.}
            \label{fig:measProc}
         \end{figure}%
         
         After a first tapping mode image providing an overview and means of orientation, a smaller measurement region is chosen in which many solitary particles are located. Secondly, a contact mode measurement is performed in this region with a scan line separation corresponding to the apparent diameter of the beads thus ensuring that most particles will only be hit once and ideally no particle is missed. This way, the nanoparticles are pushed out of the measurement region. The first peak in the recorded lateral force corresponding to the first hit of a particle is identified from this data as lateral force associated with the removal of this particle. Finally, a second tapping mode image of this particular region gives an overview of the particle removal and allows to identify the height of the plateaus underneath the particles.
         
         Basic post processing to correct image artefacts like bow and tilt is done with gwyddion~\cite{bib:Necas2012}, the data processing takes place in Mathematica~\cite{bib:Mathematica}. Following the correction of an offset in the force resulting from a tilt of the AFM tip induced by dynamic friction, the maximal peak in the lateral force is determined. Obtaining the plateau height requires more finesse. While the approximate region of the individual plateaus can be determined via inbuilt Mathematica component detection methods, the actual plateau height can be determined accurately from a fit. Since we studied the regime of small heights where the average plateau height stayed below $0.6\,\mathrm{nm}$, the tip broadening washed out any structure on and around the plateaus so that a parabola fit accurately reproduced their shape. The plateau height is then given by the maximum of the parabola and the corresponding error is determined from the fit errors and parameters. This procedure allows the very precise calculation of the plateau height so that heights below $0.1\,\mathrm{nm}$ can still be resolved despite the adverse signal-to-noise ratio for such small heights.
         
         In order to achieve high temporal resolution, we performed measurements every or every few days.
         Both lateral force and height were measured at predetermined times over $32$ days. The number of measured particles per measurement day ranged between $19$ and $39$ particles. For most of the measurement days there were well over $25$ particles. This ensures meaningful statistics and allows the statistically significant determination of both the lateral force and the plateau height.
         
         {\bf\noindent Storage:} Between measurements the sample was stored in a wet nitrogen environment inside a desiccator. In order to flush out the ambient air, it was pumped out to create a vacuum. After closing the valve connected to the vacuum pump, the nitrogen gas was slowly released into the desiccator after passing through an ultra pure Millipore water~\cite{bib:millipore} reservoir thus creating a humid environment inside the vessel. The valve was closed once first signs of condensation are visible on the inside of the desiccator. This process was repeated three times in order to increase the purity of the gas mixture. \\
         Both storage and measurements were carried out at an ambient temperature of approximately $20\,^\circ\mathrm{C}$. The sample was exposed to ambient air only for the comparatively short period of one hour during the measurements.
      \section{Results}
         \label{seq:resultsMeas2}
         \begin{figure}[htbp]
            \centering
            \includegraphics[width=.48\textwidth]{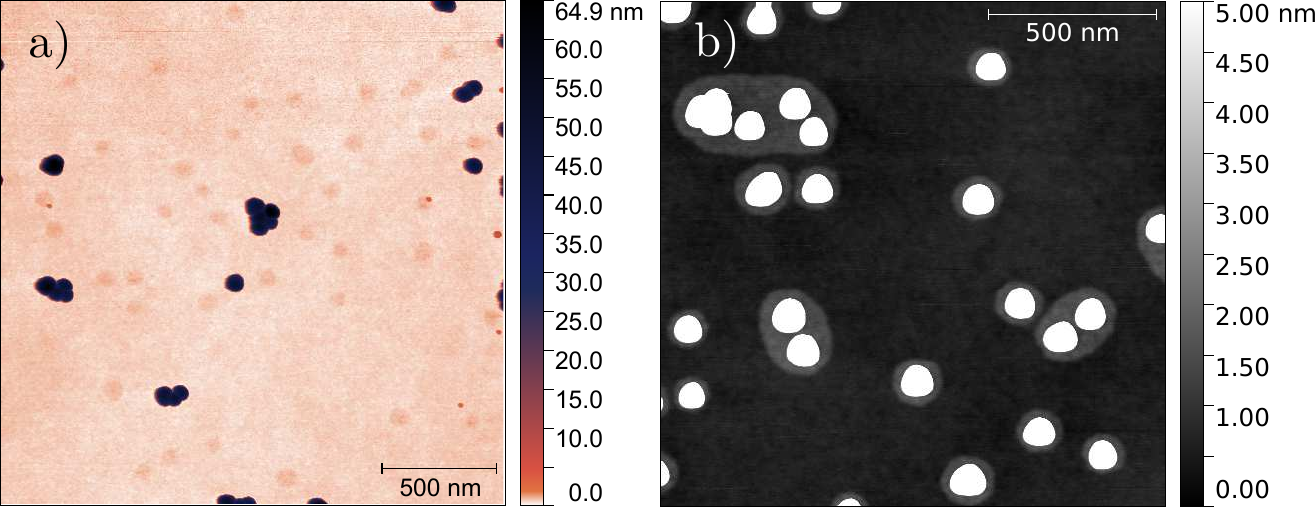}
            \caption{a) Plateaus underneath the nanoparticles visible after the removal of the particles with an AFM cantilever tip and b) large plateaus around particles on a sample stored in ambient air for 8 weeks. The height scale in the second image was cut at $5\,\mathrm{nm}$ in order to reveal the plateaus. The nanoparticles with an average height of $50\,\mathrm{nm}$ therefore appear white.}
            \label{fig:plateausUnderneath}
         \end{figure}%
         After the particle removal via the described method, plateaus become visible in the AFM image in the places of the particles, see Fig.~\ref{fig:plateausUnderneath}a).
         \begin{figure}[htbp]
            \centering
            \includegraphics[width=.45\textwidth]{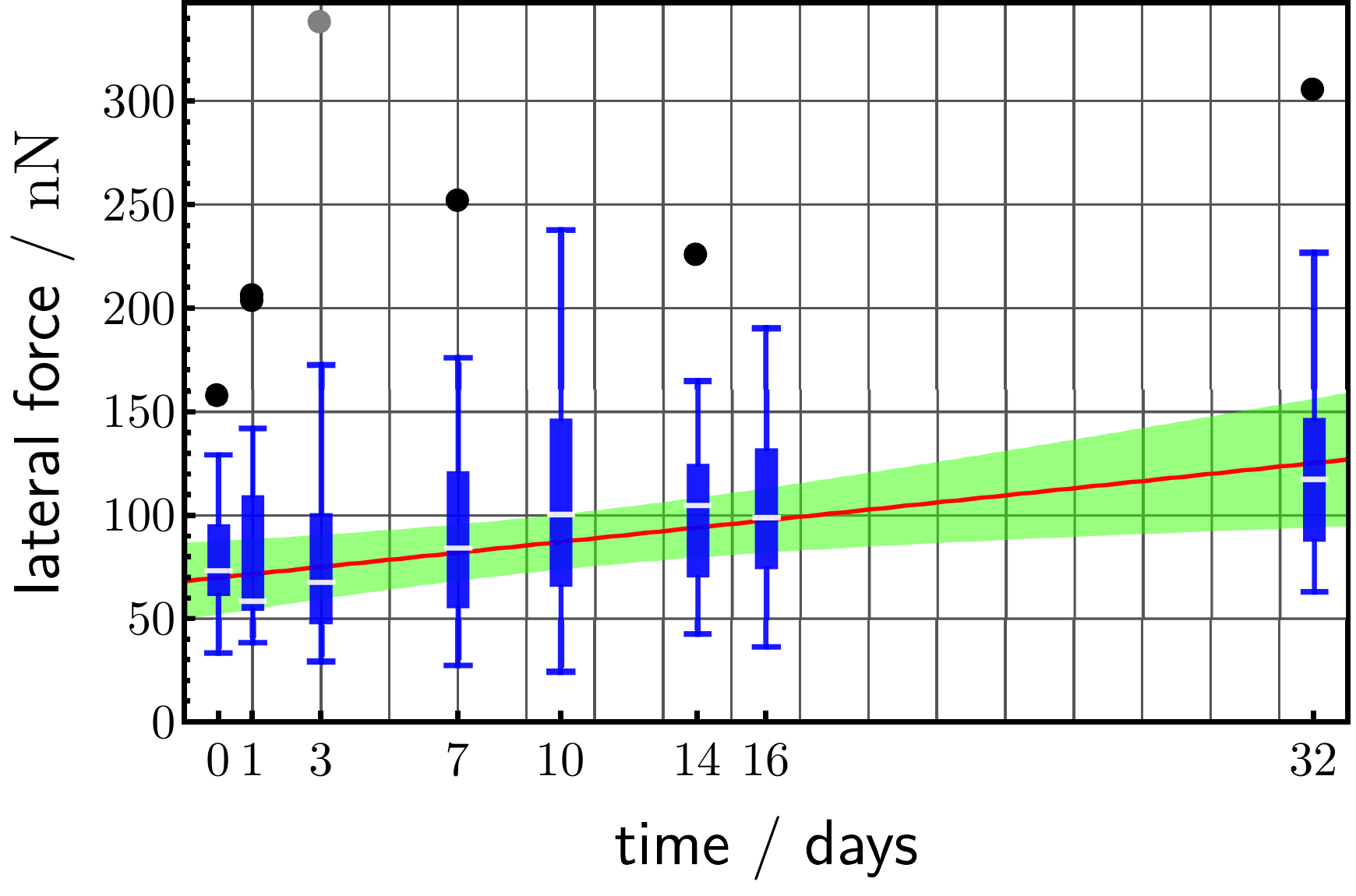}
            \caption{Time development of the lateral force with $99\,\%$ single prediction bands. In absence of a suitable model, a line was fitted as the easiest increasing function to reveal the increase.}
            \label{fig:lateralForceMeas2}
         \end{figure}%
         \begin{figure}[htbp]
            \centering
            \includegraphics[width=.48\textwidth]{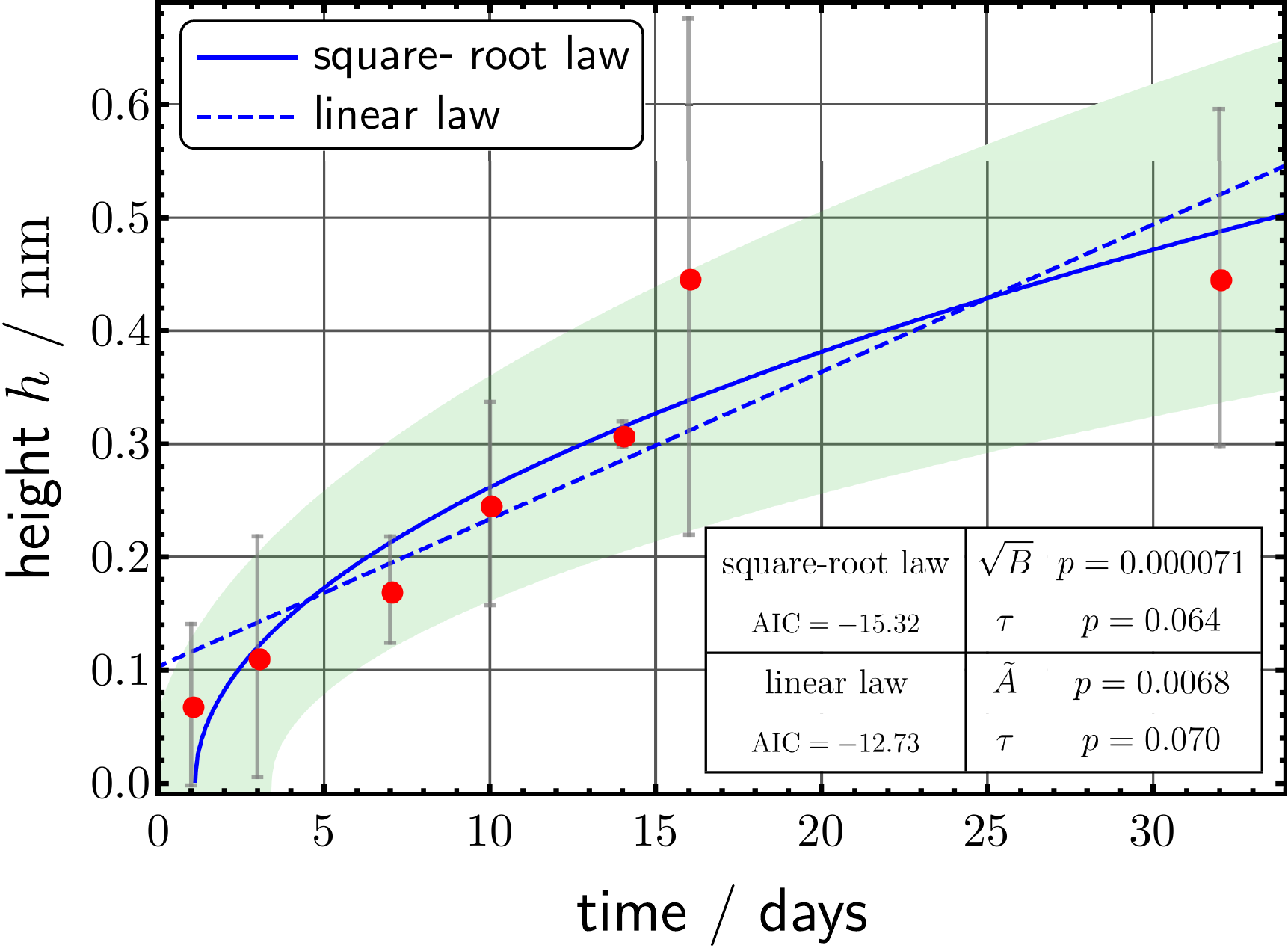}
            \caption[Square-root fit and linear fit to the plateau height depending on the storage time in wet nitrogen]
                    {Square-root fit and linear fit to the plateau height depending on the storage time in wet nitrogen. The shaded region is bounded by a square-root fit through the lower and upper limit given by the error bars representing the standard deviation. Lines denote a fit through the data points.
                    This results in \\
                    $h_\mathrm{sqrt}(t) = \sqrt{(0.088\pm0.008)\mathrm{\frac{nm^2}{day}}\cdot t + (-1.2\pm0.5)\mathrm{nm^2}}$ \\
                    for the square-root law and in \\
                    $h_\mathrm{lin}(t) = (0.10\pm0.05)\mathrm{nm} + (0.013\pm0.003)\mathrm{\frac{nm}{day}}\cdot t$ \\
                    for the linear law.
                    As seen from the inset, both p-value and Akaike information criterion (AIC)~\cite{bib:Akaike} are in favour of the square-root law.
            }
            \label{fig:heightMeas2}
         \end{figure}%
         
         Figs. \ref{fig:lateralForceMeas2} and \ref{fig:heightMeas2} show the development of lateral force and height with time, respectively. The increase in lateral force at the release of the particle may not be significant at first sight, but the fit of the lateral force to a linear model with single prediction bands proves that the increase is statistically significant. The large variation of the force values stems from the probabilistic scanning method \cite{bib:Geiger}. The single prediction bands mark the region of $99\,\%$ confidence level. We chose a linear model as the most simple monotonically increasing function. Since the upper boundary of the single prediction band on day $0$ is lower than the lower boundary of the single prediction band on day $32$, one can conclude with $99\,\%$ certainty that there is an increase in lateral force during this time span. We propose possible models for the functional dependence of the force on height (and time) below.
         
         The plateau height also shows a significant increase, which we compare to the Deal-Grove model for the height $h$ of the oxide layer after time $t$. It has two extreme cases:
         a square-root law ${h(t) = \sqrt{B(t+\tau)}}$ for large time when diffusion dominates, and a linear law ${h(t) = \tilde A(t+\tau)}$ for short times when thermal activation plays the main role. $\tilde A$ and $B$ are coefficients characterising the velocity of the reaction depending on the effective diffusion constant and the partial pressure, and $\tau$ accounts for an initial oxide layer. For further details see the supplementary material.
         
         In Fig.~\ref{fig:heightMeas2} we compare the two models.
         The linear law is unlikely since it predicts the existence of an oxide layer in place of the plateaus before the sample preparation.
         Both the p-values and the Akaike criterion obtained from the fit show that the square-root law agrees much better with the measurement results.
         Using this model, we determined the constant of the Deal-Grove model ${B = (3.2 \pm 0.6)\cdot 10^{-10}\,\mathrm{(\mu m)^2\,h^{-1}}}$.
         As discussed in the supplementary material, this value agrees with literature results.
         This agreement of the Deal-Grove model with the development of the plateau height clearly is a strong indication that the growing material is silicon oxide. The prevalence of the square-root law over the linear contribution indicates that the process is dominated by diffusion and that the thermal activation time is less relevant which is plausible in this temperature and time regime.

\begin{figure}[htbp]
            \centering
            \includegraphics[width=.45\textwidth]{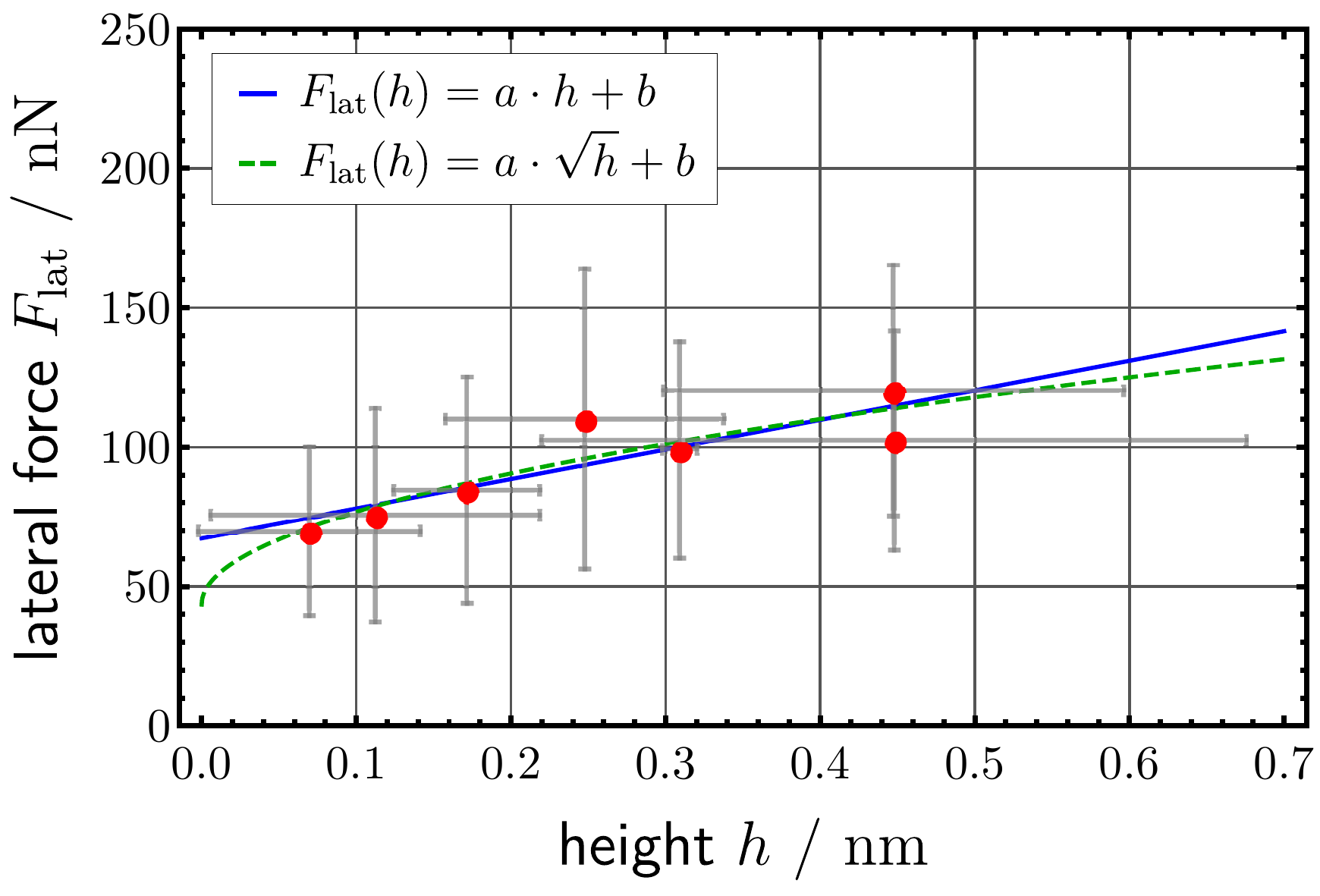}
            \caption[Correlation between lateral force and plateau height]
                    {Correlation between force and plateau height.
                     The two suggested models yield the following fit results: the linear law corresponding to the proportionality of the lateral force to the contact area between plateau and particle reads \\
                     $F_\mathrm{lat, a} = (67\pm8)\mathrm{nN} + (106\pm27)\mathrm{nN\,nm^{-1}}\cdot h$, \\
                     whereas the proportionality to the circumference results in \\
                     $F_\mathrm{lat, c}=(43.5\pm11.8)\mathrm{nN} + (105\pm23)\mathrm{nN\,nm^{-1/2}}\cdot\sqrt{h}$.
                    }
            \label{fig:forceHeight}
         \end{figure}

         {\bf\noindent Correlation of Lateral Force and Plateau Height:} Fig.~\ref{fig:forceHeight} shows the lateral force corresponding to a particular plateau height. 
         A clearly increasing trend is visible, for which we propose two possible functional dependences below.
         
         To devise a relation between the plateau height and the measured lateral force upon particle release, we assume that the lateral force is proportional to the interaction forces between particle and contact surface \cite{bib:Geiger}. Picturing a spherical particle lying in a spherical shell of silicon oxide growing around the particle, we suggest two intuitive options:
         1) the lateral force is proportional to the contact area between particle and spherical shell or 2) it is proportional to the circumference of the shell, as for a droplet.
         Additionally, one may expect an offset in the force due to the formation of a water meniscus around the particles~\cite{bib:ranade}.
         
         Geometrical considerations lead to a linear dependence of the lateral force on the plateau height for option one, whereas option two results in a dependence on the square root of the height. The two models can therefore be written as
         ${F_\mathrm{lat,a}(h) = a \cdot h + b}$ and ${F_\mathrm{lat,c}(h) = a \cdot \sqrt{h} + b}$, respectively.
         While both models agree with the data, the fit results do not show a statistically significant distinction of one over the other.
         
         Since we attribute the increase in plateau height to the growth of an oxide layer around the particle and there appears to be a strong correlation between lateral force and height, we conclude that both effects are caused by the same mechanism, namely in this case the oxidation process of silicon.
         
      {\bf\noindent Plateau Formation Mechanism:} 
      Apart from agreeing with the Deal-Grove model we justify our claim that the forming material is silicon oxide with additional observations.
      The plateaus are robust against multiple scans in contact mode with increased normal force, i.e. the adhesion or cohesion is strong enough to prevent the layers from being scratched off with a silica tip.
      This also rules out that these layers arise from the agglomeration of macromolecules stemming from a contamination. Such molecules should have already been visible in the AFM scans directly after the preparation, yet our samples appear clean. 
      This is supported by our observations made under equal storage conditions but in a dry nitrogen environment with no oxidising substances present, in which the plateau formation process is suppressed. Hence, oxide growth is a viable explanation.
      
      A second qualitative measurement series not reported here with storage in a pure oxygen environment also showed the growth of plateaus, albeit at a slower rate. This is consistent with a lower activation energy for the diffusion process in a wet environment rather than pure oxygen gas leading to a larger growth constant $B$ in the former case \cite{bib:Deal}. The formation of plateaus in this environment marks oxygen as vital ingredient in the process and makes a contribution of nitrogen unlikely.
      
\begin{figure}[htbp]
         \centering
         \includegraphics[width=.38\textwidth]{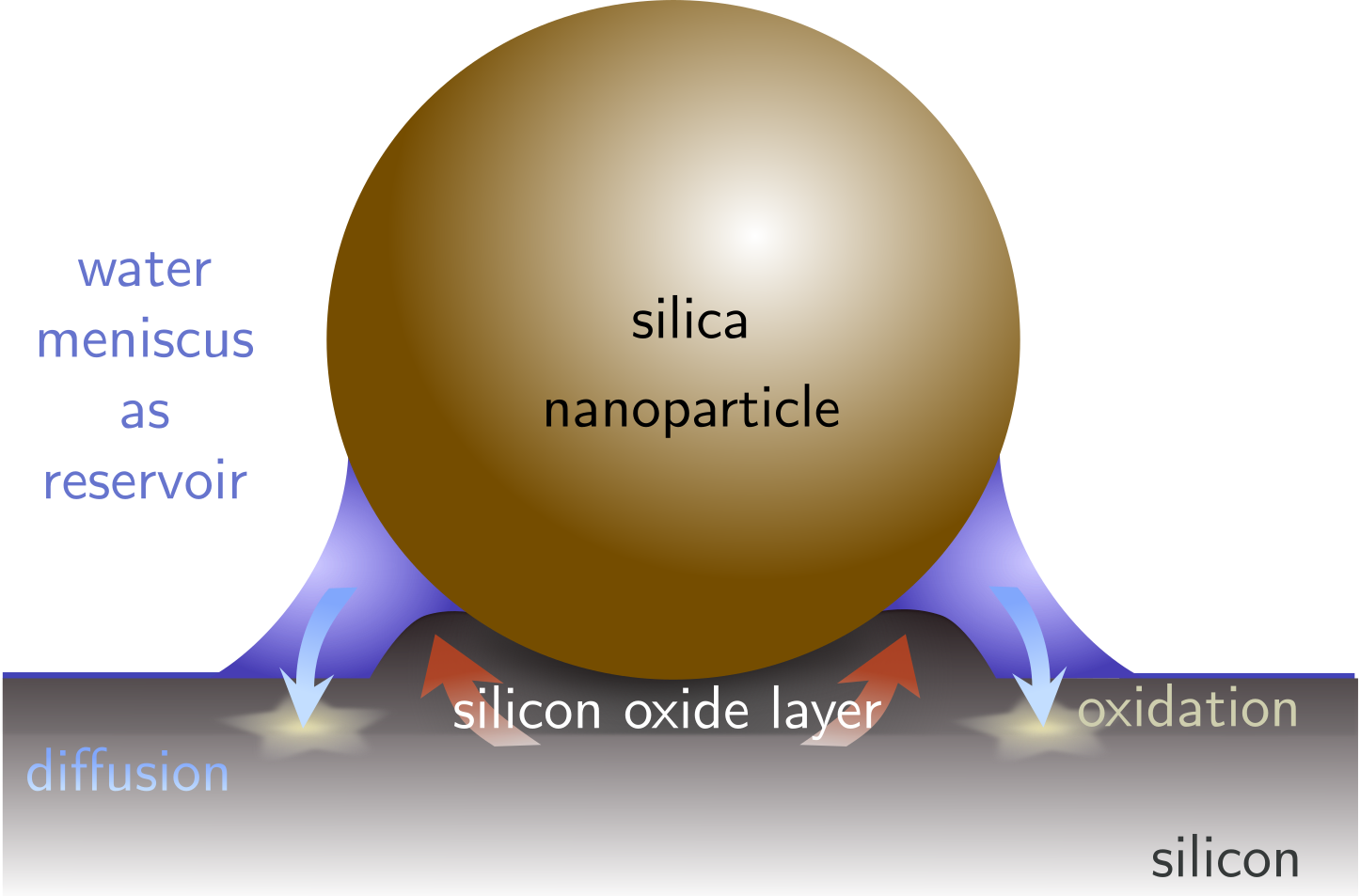}
         \caption{Oxidation process: condensation around the nanoparticles leads to the formation of a meniscus acting as water reservoir. Water diffusing through the oxide layer reaching the silicon bulk material leads to the gradual build-up of an oxide layer around and beneath the nanoparticles.}
         \label{fig:oxidationSketch}
      \end{figure}

      Concluding that the plateaus consist of oxide, we propose the following explanation for the localisation of these structures around the particles. Acting as condensation seeds, a water meniscus naturally forms as a ring around them between particle and silicon surface. The locally higher concentration of oxidising substance leads to a faster growth of oxide beneath the particles, see Fig.~\ref{fig:oxidationSketch}. The meniscus effectively acts as water reservoir for the diffusion of the water molecules through existing oxide layers to the native silicon, where an oxidation reaction takes place. Since the oxide grows into the bulk and since both $\mathrm{SiO}$ and $\mathrm{SiO}_2$ will be created, the large resulting tensions will result in the build-up of an amorphous silicon oxide layer and will lift up the surface around the silica particles. \\
      As the nanoparticles also consist of silica, the formation of chemical bonds between particles and surface is likely, which leads to an increase in adhesion force in accordance with recent studies of frictional aging~\cite{bib:Li,bib:Tian2017,bib:Tian2018,bib:Tian2018PhysChem}. \\
      Once the plateaus have reached a certain height, the water meniscus will form at the flank of the plateau rather than between particle and surface leading to a growth of the plateaus in area rather than height. This is visible on samples stored in humid environments for many weeks as the one in Fig.~\ref{fig:plateausUnderneath}b). On such samples, the plateaus reach heights on the order of $1$-$2\,\mathrm{nm}$ and extend over multiple hundred square micrometer.
      
      While we observe a differential increase of the plateau height around the particles, we also expect oxidation to take place on the rest of the sample, but at a slower rate. According to \cite{bib:opitz} silicon samples handled in ambient air are covered by a thin water film of approximately $1\,\mathrm{nm}$ thickness. Being much thinner, the resulting water concentration in the surface layers leads to a slower growth as $B$ is proportional to the concentration \cite{bib:dealGrove}. Consequently, we can detect the oxide height around the particles only in contrast to their neighbourhood, which is expected to grow much more slowly.
      This inhomogeneous growth leads to a considerable increase in surface roughness, which persists even after the removal of the nanoparticles.
      
\begin{figure}[htbp]
         \centering
         \includegraphics[width=.3\textwidth]{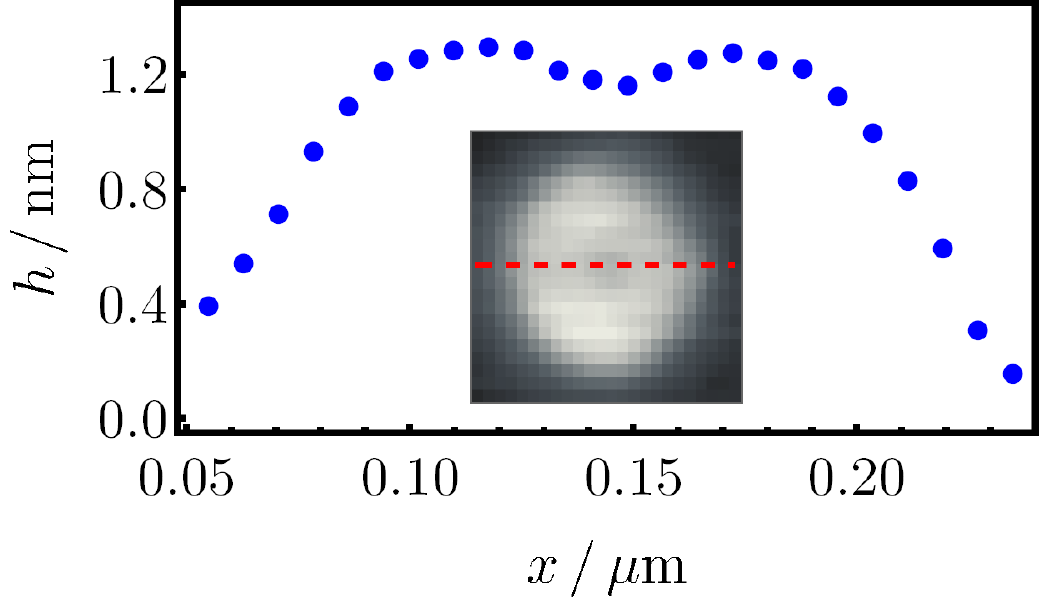}
         \caption{
            AFM tip broadened profile of a plateau averaged over $5$ profiles on the same measurement day.
            A small dip in the centre of the plateau corresponding to the previous position of the particle is revealed.
            }
         \label{fig:dip}
      \end{figure}

      Due to the shape of the water meniscus, we expect the oxide to grow around the particle forming a spherical shell and we thus anticipate to see a dip in the middle of the plateaus. For small layer heights this is difficult to observe since the used AFM tips do not allow to resolve such fine spherical hollows very well. However, for larger heights it is indeed possible to see the aforementioned dip as shown in Fig.~\ref{fig:dip}.
      
   \section{Conclusion}
      We have shown the growth of oxide structures (plateaus) beneath and around nanoparticles, as well as investigated the increase in lateral removal force depending on the storage time in a wet nitrogen environment. Moreover, we have demonstrated a correlation between lateral force and plateau height indicating that both processes are governed by the same mechanism. Since the plateaus likely grow due to an oxidation process, this points to a chemical reaction also being responsible for the increased adhesion of the particle.
      
      Furthermore, we have shown that our results for the oxide growth are in accordance with the Deal-Grove model. Their original model described the growth of extended oxide layers with a homogeneous distribution of oxidising substance on the surface. Remarkably, the model still seems to apply here despite the low temperatures and the locally enhanced concentration.
      
      While we observe a layer growth of only about $0.6\,\mathrm{nm}$ in $32$ days, we expect the growth to continue and have observed heights of $1$-$2\,\mathrm{nm}$ on samples stored for $8$ weeks. A plateau thickness in the nanometer range is already likely to lead to diffraction and scattering losses in photolithography applications, where the wavelength is on the order of $10\,\mathrm{nm}$. Furthermore, the plateaus increase the surface roughness considerably. The chemical structure is altered locally due to the presence of the nanoparticles leading to the irreparable damage of the surface.
      
      Therefore, this study shows that in order to maintain a smooth silicon surface it is necessary to keep them clean from dust and other particle sources and store them in the absence of oxidising substances, such as water vapour and oxygen.
      
   \section{Acknowledgements}
      C.W. is grateful for the financial support received by the German Academic Scholarship Foundation. Furthermore, we would like to thank Ivana Sfarcic who assisted in the chemical lab, and the other members of the Institute for Experimental Physics at Ulm University. We are grateful to Kay-Eberhard Gottschalk (Institute for Experimental Physics, Ulm University) for fruitful discussions.
   

   \appendix
   
   \clearpage

   \begin{center}
   \widetext
   \large\bf Deal-Grove Oxidation and Nanoparticle Adhesion -- an AFM Study \\ Supplemental Material \\[\baselineskip]
   \end{center}
   \twocolumngrid
   
   \setcounter{equation}{0}
   \setcounter{figure}{0}
   \setcounter{table}{0}
   \setcounter{page}{1}
   
   \section{The Deal-Grove Model}
      The Deal-Grove model \cite{bib:dealGrove} describes the thermally activated, diffusion limited growth of silicon oxide on a silicon surface by an oxidising gas or vapour. For the growth of an oxide layer of thickness $h$ Deal and Grove found \cite{bib:dealGrove}
      \begin{align}
         h(t) & = \frac{A}{2} \left(\sqrt{1+\frac{4B}{A^2}(t+\tau)} - 1\right)\text{,}
      \end{align}
      where $A$ and $B$ are coefficients characterising the velocity of the reaction depending on the effective diffusion constant and the partial pressure of the oxidising gas or vapour. $\tau$ accounts for an initial layer on the sample. This equation simplifies for different time regimes: for small times, i.e. $t+\tau \ll \frac{A^2}{2B}$, one finds
      \begin{align}
         h(t) & = \frac{B}{A} (t+\tau)\text{,}
      \end{align}
      whereas for large times $t+\tau \ll \frac{A^2}{2B}$ as in this experiment the diffusion process dominates over the thermal activation and one recovers
      \begin{align}
         h(t) & = \sqrt{B(t+\tau)}.
      \end{align}
   \section{Comparison to Different Temperature Regimes}
      Originally, Deal and Grove observed a rapid oxidation growth process for wet oxygen at temperatures in a range between $1100\,\mathrm{K}$ and $1500\,\mathrm{K}$. This raises the question, whether the observations can be attributed to the same mechanism. However, the following comparison of the temperature dependent coefficient $B$ from our measurement data and the respective coefficient extracted from the data given by Deal and Grove shows that they are of the same order of magnitude.
      
      \begin{figure}[htbp]
         \centering
         \subfigure[\ Fit to the values of $B$ obtained by Deal and Grove \cite{bib:dealGrove}]{
            \includegraphics[width=.45\textwidth]{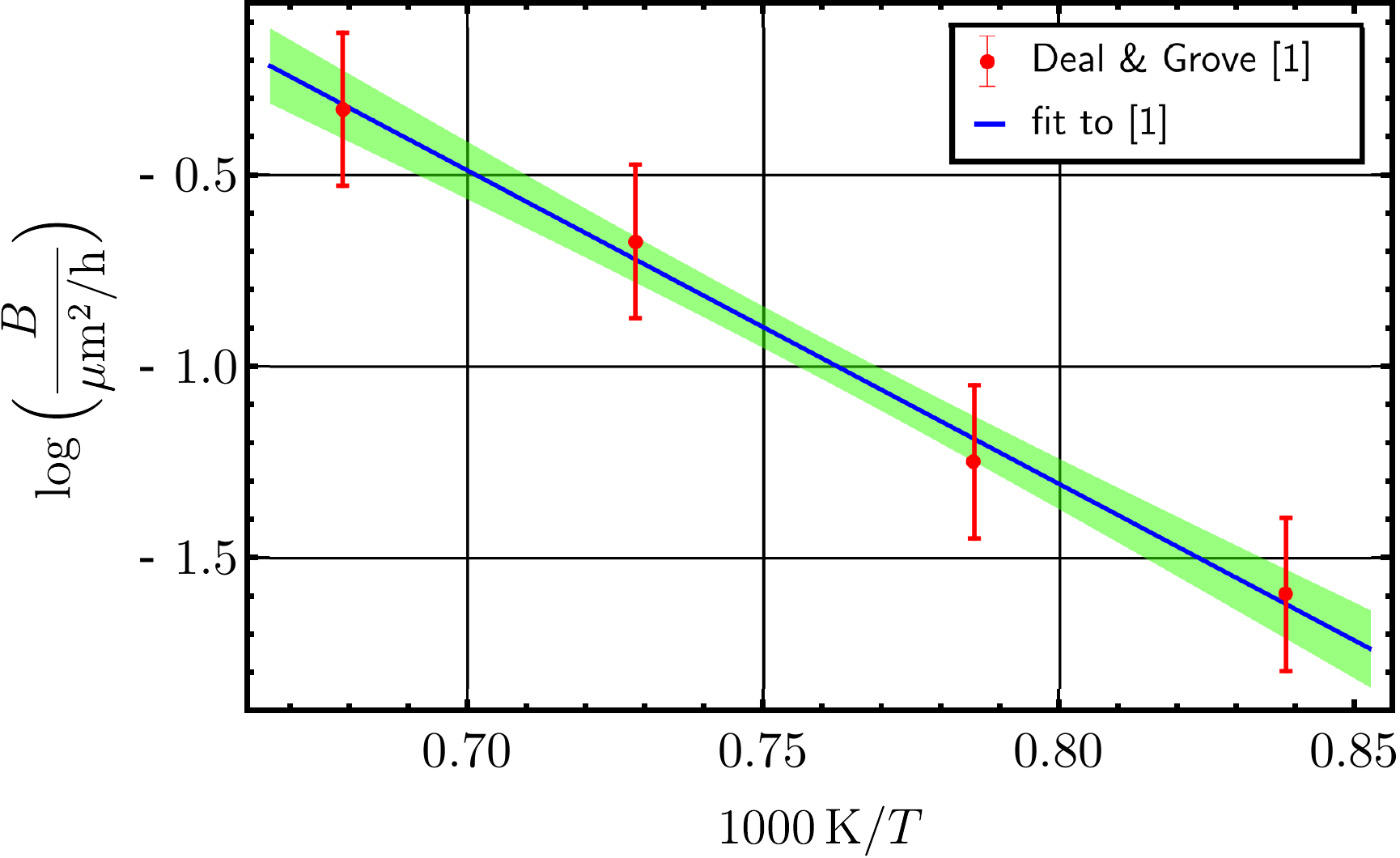}
            \label{fig:DealGrovesData}
         }
         \subfigure[\ Fit to the values of $B$ obtained by Deal and Grove \cite{bib:dealGrove} with the result for $B$ of this study]{
            \includegraphics[width=.45\textwidth]{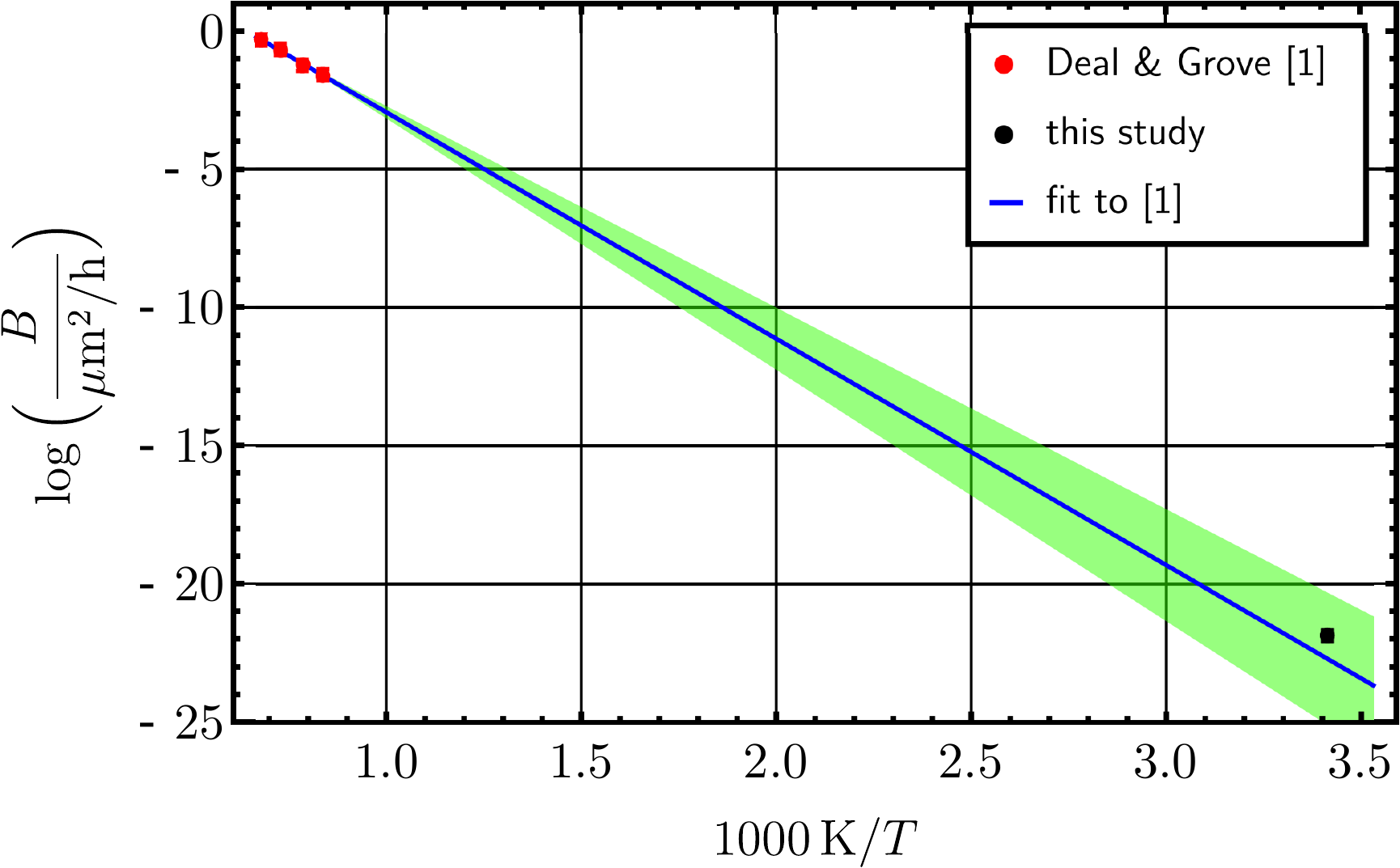}
            \label{fig:DealGrovesDataPlusOwn}
         }
         \caption{Temperature dependence of the constant $B$ from the Deal-Grove model with $80\,\%$ single prediction bands. Our data point lies within the confidence interval obtained from a fit to Deal and Grove's data.}
      \end{figure}%
      Deal and Grove determined the constant $B$ in their paper for wet oxygen at four different temperatures. They state the dependence of $B$ as
      \begin{align}
         B & = \frac{2 D_\mathrm{eff} C^*}{N_1}\text{,}
      \end{align}
      where $D_\mathrm{eff}$ is the effective diffusion constant, $C^*$ is the equilibrium concentration of the oxidant in the oxide and $N_1$ is the number of molecules of the oxidant inside a volume unit of the oxide layer \cite{bib:dealGrove}. According to Henry's law, the diffusion constant $C^*$ is proportional to the partial pressure of the oxidant. Doremus \cite{bib:Doremus} gives the temperature dependence of $B$ as
      \begin{align}
         B & = B_0 \, \mathrm{exp}\left(-\frac{Q}{R T}\right)\text{,}
      \end{align}
      with the activation energy $Q$ of the diffusion process. Fitting this temperature dependence in the form of
      \begin{align}
         \log \frac{B}{\mathrm{\mu m^2 h^{-1}}} & = a - b\cdot \frac{1}{T}.
      \end{align}
      to the data determined by Deal and Grove and comparing it to the value of $B$ obtained in this study at $293\,\mathrm{K}$, we find a very good agreement. Fig.~\ref{fig:DealGrovesData} shows the fit containing the original data of Deal and Grove, while Fig.~\ref{fig:DealGrovesDataPlusOwn} shows a larger temperature range including our value of $B$ at room temperature. The error bars are too small to be visible in this scaling. The value of $B$ obtained in this study lies close to the line given by the fit through the data of Deal and Grove. Thus, it seems plausible that the same growth process takes place at room temperature, although more data points at different temperatures would be preferable to check this statement. In addition, we expect the same process to occur since there are no phase transitions of silicon within this temperature range. \\
      The small discrepancy between the fit to the data by Deal and Grove and the result from this study may stem from different types of oxidants. Deal and Grove used wet oxygen, whereas wet nitrogen was used for this measurement series. In addition, one needs to take into account that a different partial pressure of the oxidant might influence the temperature dependence of $B$. Nevertheless, the good agreement speaks in favour of oxidation taking place on our sample leading to the plateau growth in our case.
   
   %
   
\end{document}